\documentclass[prb,showpacs,twocolumn,amsmath,amssymb,floatfix]{revtex4-1}
\usepackage{graphicx}
\usepackage{bm}
\usepackage{bbm}
\usepackage{color}
\usepackage{footnote}
\usepackage{epstopdf}
\usepackage{hyperref}
\usepackage{appendix}
\usepackage[applemac]{inputenc}
\begin{document}
\title{Comparison of methods for computing the exchange energy in laterally coupled quantum dots}
\author{Jorge Cayao}
\affiliation{Faculty of Informatics, Physics and Mathematicas, Comenius University, Bratislava 842 48, Slovakia}
\date{September 17, 2010} 
\begin{abstract}
We  calculate the exchange energy in two dimensional laterally coupled quantum dots using  Heitler-London, 
Hund-Mullikan and variational methods.  We  assess the quality of these approximations in zero and finite magnetic fields comparing against numerically exact results.
We find  that surprisingly, the Hund-Mullikan method does not offer any significant improvement over the much simpler Heitler-London method, whether at large or small interdot distances. 
Contrary to that, our variational ansatz proves substantially better.  In a single dot at finite magnetic field, all approximate methods fail.
This reflects the qualitative change of the single electron ground state from non-degenerate (harmonic oscillator) to highly degenerate (Landau level).
However, we find that the magnetically induced failure does not occur in the most important, double-dot, regime.
\end{abstract}
\maketitle
\section{Introduction}
In the last decade, a great interest in quantum dots has
aroused, due to their potential use as hardware for a scalable quantum computer \cite{qcomp, inom, quantumcomputer, teoria4}.
 In this implementation, the electron spin in quantum dots is used as the basic unit of information (qubit) \cite{twoelectronqdmol}. 
To perform any quantum algorithm, 1-qubit and 2-qubit gates are sufficient.
 Any 2-qubit gate can be achieved through single qubit rotations and an adequate switching of the exchange energy, which parametrizes the spin
coupling in the Heisenberg spin exchange  Hamiltonian \cite{quantumdots, teoria1}.
The knowledge of the exact value of the exchange energy is important as it determines the time of the the $\sqrt{SWAP}$, a fundamental 2-qubit gate \cite{qcomp}. 
The properties of the exchange energy in lateral dots  have been investigated by a variety of methods:
Heitler-London \cite{PhysRevB.76.125323, PhysRevB.59.2070, quantumcomputer, inom, couledquantumdots}, Hund-Mulliken \cite{PhysRevB.76.125323, quantumcomputer}, 
Molecular Orbital \cite{quantumcomputer}, Variational \cite{Variational}, Configuration Interaction \cite{PhysRevB.73.155301, ci, config3}, 
Hartree \cite{ exacttreat}, Hartree-Fock \cite{quantumcomputer,  exacttreat}, Hubbard model \cite{quantumcomputer}, quantum Monte Carlo \cite{monte} and local 
spin density approximation \cite{monte, lsd}.

This work compares several standard methods (Heitler-London, Molecular Orbital, Hund-Mulliken, Variational and the
exact Configuration Interaction) to compute the exchange energy in quantum dots. We find that, with the exception of our Variational ansatz,
 the other standardly used extensions of the Heitler-London method do not in fact offer any real improvement. We revisit the failure of the Heitler-London method  in a
finite magnetic field \cite{couledquantumdots}. We explain the failure as due to the qualitative change of the single electron ground state
from non- to highly- degenerate. As our most important result, we find that  contrary to the singlet dot case, the Heitler-London method (and thus all the considered approximate methods) does not suffer 
the failure in the magnetic field in the double dot regime.

The article is organized as follows. In Sec.~\ref{sec.II} we define  the double dot model, and  present the Fock-Darwin states.
 In Sec.~\ref{sec.III}  we define the approximate methods that we compare in further.
In Sec.~\ref{sec.IV}, we present the comparison, varying the magnetic field and the interdot distance.
\section{Model}
\label{sec.II}
We assume to have two electrons in a  harmonic electrostatic potential with one (quantum helium), 
or two symmetrical (quantum hydrogen) minima \cite{exacttreat, fromHetoH2, Variational, confinementpotential, exactsolutionqds}. 
We consider the electron to be two-dimensional, as appropriate for electrically
defined lateral semiconductor quantum dots \cite{inom, teoria4, teoria1, schranaliticsolut} in GaAs/AlGaAs heterostructure. 
We describe the electrons using the single band effective mass approximation \cite{fabian}.
A constant magnetic field is applied along the growth direction (being here along the $z$-axis) \cite{schranaliticsolut, magfielddepoflow-energy}.

For the quantum hydrogen (\emph{double dot}) the Hamiltonian of the $i$-th electron is given by \cite{fromHetoH2} 
\begin{equation}\label{eq:hydrohami}
 H_{d}^{(i)}=\frac{{\bf P}_{i}^{2}}{2m}+\frac{1}{2}m\omega^{2}{\rm min}\lbrace({\bf r}_{i}-{\bf d})^{2},({\bf r}_{i}+{\bf d})^{2}\rbrace,
\end{equation}
where the kinetic momentum  ${\bf P}={\bf p}+e{\bf A}$ is expressed using the canonical momentum
 ${\bf p}=-{\rm i}\hbar {\bf \nabla}$,  and the vector potential ${\bf A}=-\frac{1}{2}{\bf R}\times{\bf B}$, projected to the two dimensional plane.
 The magnetic field ${\bf B}=(0,0,B)$, and the position vector ${\bf R}=({\bf r},z)$, ${\bf r}=(x,y)$. The positron
 elementary charge is $e$, the effective mass of the electron is $m$, and $\hbar\omega$ is the confinement energy \cite{confinementpotential}. 
The vector {\bf d} defines the main dot axis with respect to the
crystallographic axes.

 The quantum helium (\emph{single dot}) can be described setting the interdot distance to zero \cite{fromHetoH2}, resulting in
the Hamiltonian, 
\begin{equation}\label{eq:hhelium}
 H_{d=0}^{(i)}=\frac{1}{2}\frac{{\bf P}_{i}^{2}}{m}+\frac{1}{2}m\omega^{2}{\bf r}_{i}^{2}\,.
\end{equation}
For a system of two electrons the Hamiltonian is
\begin{equation}\label{eq:hamilto}
 H=H_{d}^{(1)}+H_{d}^{(2)}+H_{C}\,,
\end{equation}
where $H_{C}$ is the Coulomb interaction Hamiltonian between the two electrons, 
\begin{equation}
 H_{C}=\frac{e^{2}}{4\pi \epsilon_{0}\epsilon_{r}}\frac{1}{|{\bf r}_{1}-{\bf r}_{2}|}\,,
\end{equation}
where $\epsilon_{0}$ and $\epsilon_{r}$ are the vacuum and material dielectric constants, respectively.
We neglect the Zeeman and spin-orbit interactions, as they are substantially smaller than the above terms\cite{spinorb}.

In the numerical computations, we use the parameters of GaAs: $\epsilon_{r} = 12.9$,  $m=0.067 m_{e}$ ($m_{e}$ is the free electron mass), take 
the confinement energy $\hbar \omega = 1$ meV \cite{data}, and place the dot such that ${\bf d}=(d,0)$, that is along the x-crystallographic axis. 

\subsection{Fock-Darwin states}
The eigenfunctions of the single dot Hamiltonian, Eq.~(\ref{eq:hhelium}), are the Fock-Darwin states \cite{inom, schranaliticsolut},
\begin{equation}\label{eq:fock}
 \psi_{n\ell}^{1e}({\bf r})=C_{n\ell}r^{|\ell|}{\rm e}^{-r^{2}/(2l_{B}^{2})}L_{n}^{|\ell|}\left(\frac{r^{2}}{l_{B}^{2}}\right){\rm e}^{{\rm i}\ell\varphi}\,,
\end{equation}
Here $C_{n\ell}$ is the normalization constant,  $L_{n}^{\ell}$ are Laguerre polynomials, and $n$,  $\ell$ are  
principal and orbital quantum numbers, respectively. The right hand side of Eq.~(\ref{eq:fock}) is expressed in polar coordinates ($r, \varphi$).
The corresponding energies read, 
\begin{equation}\label{eq:40}
 E_{n,\ell}=\frac{\hbar^{2}}{l_{B}^{2}m}(2n+|\ell|+1)+\frac{eB\hbar}{2m}\ell\,, 
\end{equation}
where,
\begin{equation}\label{eq:effe}
 l_{B}=\left(\frac{e^{2}B^{2}}{4\hbar^{2}}+\frac{m^{2}\omega^{2}}{\hbar^{2}}\right)^{-1/4}\,,
\end{equation}
is the effective confinement length.
\subsection{Exchange Energy}
Neglecting the Coulomb interaction, one can write the two-electron wave-function using single electron eigenstates. 
If the state is separable to the spinor and orbital parts, to have an
antisymmetric function, orbital part must be symmetric and spinor antisymmetric, or vice versa.
         At low-temperature the relevant two electron Hilbert space can be restricted to comprise the two lowest orbital eigenstates of the Hamiltonian 
in Eq.~(\ref{eq:hamilto}). Such restricted subspace is described by an effective Hamiltonian, 
\begin{equation}
 \label{eq.heise}
H=\frac{J}{\hbar^{2}} {\bf S}_{1}\cdot {\bf S}_{2}\,,
\end{equation}
where ${\bf S}_{1,2}=\hbar \boldsymbol{\sigma}_{1,2}/2$ are spin operators and the only parameter is $J$,  the exchange energy.
By this construction, the exchange energy is defined as the difference between the energy of the triplet and the singlet state \cite{magte,exchange2},
\begin{equation}\label{eq:exchangee}
 J=\langle \Psi_{T}|H|\Psi_{T} \rangle-\langle \Psi_{S}|H|\Psi_{S} \rangle\,,
\end{equation}
where $\Psi_{S}$ and $\Psi_{T}$ are the spin singlet and triplet two-electron wave functions, respectively.
These wave-functions are constructed from the single electron states (Fock-Darwin) appropriately, according to the specific method.
After that, we evaluate integrals in Eq.~(\ref{eq:exchangee}) numerically, to obtain the exchange energy.
\section{Methods}
\label{sec.III}
In this section we define four approximative methods to compute the exchange energy for a system  of  two electrons 
in a coupled double dot system. They differ in the way how the two electron wave-functions, in Eq.~(\ref{eq:exchangee}), are constructed.
\subsection{Heitler-London method}
The Heitler-London approximation is the simplest method to calculate the exchange energy in a two electron dot \cite{exchange2HL, quantumcomputer}.
 It employs the lowest Fock-Darwin state ($n=\ell=0$), displaced to the potential minima positions $\pm {\bf d}$ \cite{couledquantumdots,inom},
\begin{equation}\label{eq:103}
 \psi_{L/R}({\bf r})=\frac{1}{l_{B}\sqrt{\pi}} \exp\left(-\frac{[(x\pm d)^{2}+y^{2}]}{2l_{B}^{2}}\pm \frac{{\rm i} edBy}{2\hbar}\right)\,.
\end{equation}
From these one constructs the two-electron singlet and triplet states as, 
\begin{equation}\label{eq:heitler}
 \Psi_{S/T}=C_{S/T}[\psi_{L}({\bf r}_{1})\psi_{R}({\bf r}_{2})\pm \psi_{R}({\bf r}_{1})\psi_{L}({\bf r}_{2})]\chi_{S/T}\,,
\end{equation}
where the singlet $\chi_{S}=\left(1/\sqrt{2}\right)\left(|\uparrow\downarrow-\downarrow\uparrow\rangle\right)$, $\chi_{T}$ is one 
of the three triplet states, $\left(1/\sqrt{2}\right)\left(|\uparrow\downarrow+\downarrow\uparrow\rangle\right)$, $|\uparrow\uparrow\rangle$\,,
$|\downarrow\downarrow\rangle$, and $C_{S/T}$ is the normalization constant. 
\subsection{Molecular Orbital method}
The basic idea of the Molecular Orbital method is to use molecular orbitals, instead of the localized Fock-Darwin states, as the basic building blocks of the two electron
wave-functions \cite{quantumcomputer, inom}. A Molecular orbital is the single
electron eigenstate of the double dot Hamiltonian, Eq.~(\ref{eq:hydrohami}).
We take the following one-electron wave-functions as approximations to the lowest two molecular orbitals,
\begin{equation}\label{eq:mol1}
 \psi_{\pm}({\bf r})=C_{\pm}\left[\psi_{L}({\bf r})\pm \psi_{R}({\bf r})\right]\,,
\end{equation}
where $C_{\pm}$ are the normalization constants. The symmetrized combinations of the previous
 wave-functions form the following two electron states,
\begin{equation}\label{eq:molor}
 \begin{split}
 \Psi_{1}&=[\psi_{+}({\bf r}_{1})\psi_{+}({\bf r}_{2})]\chi_{S}\,,\\
 \Psi_{2}&=[\psi_{-}({\bf r}_{1})\psi_{-}({\bf r}_{2})]\chi_{S}\,,\\
\Psi_{3}&=C_{3}\left[\psi_{+}({\bf r}_{1})\psi_{-}({\bf r}_{2})+\psi_{+}({\bf r}_{2})\psi_{-}({\bf r}_{1})\right]\chi_{S}\,,\\
\Psi_{4}&=C_{4}\left[\psi_{+}({\bf r}_{1})\psi_{-}({\bf r}_{2})-\psi_{+}({\bf r}_{2})\psi_{-}({\bf r}_{1})\right]\chi_{T}\,,
\end{split}
\end{equation}
where, again, $C_{3}$ and $C_{4}$ are normalization constants.
To obtain the two electron energies, we diagonalize the Hamiltonian in Eq.~(\ref{eq:hamilto}) in the basis consisting of
 these four states.
\subsection{Hund-Mullikan Method}
In this method one expands the two electron basis used in the Heitler-London  method, Eq.~(\ref{eq:heitler}), by doubly occupied 
states \cite{couledquantumdots}, $\psi_{L}({\bf r}_{1})\psi_{L}({\bf r}_{2})$ and $\psi_{R}({\bf r}_{1})\psi_{R}({\bf r}_{2})$. 
 It is simple to check that due to the choice of the single electron molecular orbitals, Eq.~(\ref{eq:mol1}), 
such expanded basis is equivalent to the one used in the Hund-Mullikan method. It then follows that the Hund-Mullikan
approximation is equivalent to the Molecular Orbital with the single electron orbitals approximated by Eq.~(\ref{eq:mol1}).
\subsection{Variational  method}
In the variational method one makes a guess for a trial wave-function,
 which depends on variational parameters\cite{Variational} . These parameters are adjusted until
 the energy of the trial wave-function is minimal. The resulting trial wave-function and its corresponding energy 
are variational method approximations to the exact wave-function and energy.
Here we use the Heitler-London type of ansatz, Eq.~(\ref{eq:heitler}), with
\begin{equation}
\Psi_{S,T}(D)=\Psi_{S,T}(d\rightarrow D)\,.
\end{equation}
 The energy is defined as the minimum over the variational parameter $D$,
\begin{equation}\label{eq:trial}
 E_{S,T}=\min\limits_{D}\langle \Psi_{S,T}(D)|H|\Psi_{S,T}(D) \rangle\,,
\end{equation}
where $H$ is given by Eq.~(\ref{eq:hamilto}) and the minimization is done numerically.
\subsection{ Configuration Interaction Method (Exact)}
The configuration interaction is a numerically exact method \cite{config3}, in which the two electron 
Hamiltonian is diagonalized in the basis of Slater determinants constructed from
numerical single electron states in the double dot potential \cite{fabio, fabio2, config3}. 
Typically, we use 21 single electron states, resulting in the relative error for energies
of order $10^{-5}$\cite{fabio}.
\begin{figure}[h!]
 \centering
 \includegraphics[width=8cm,height=5.5cm]{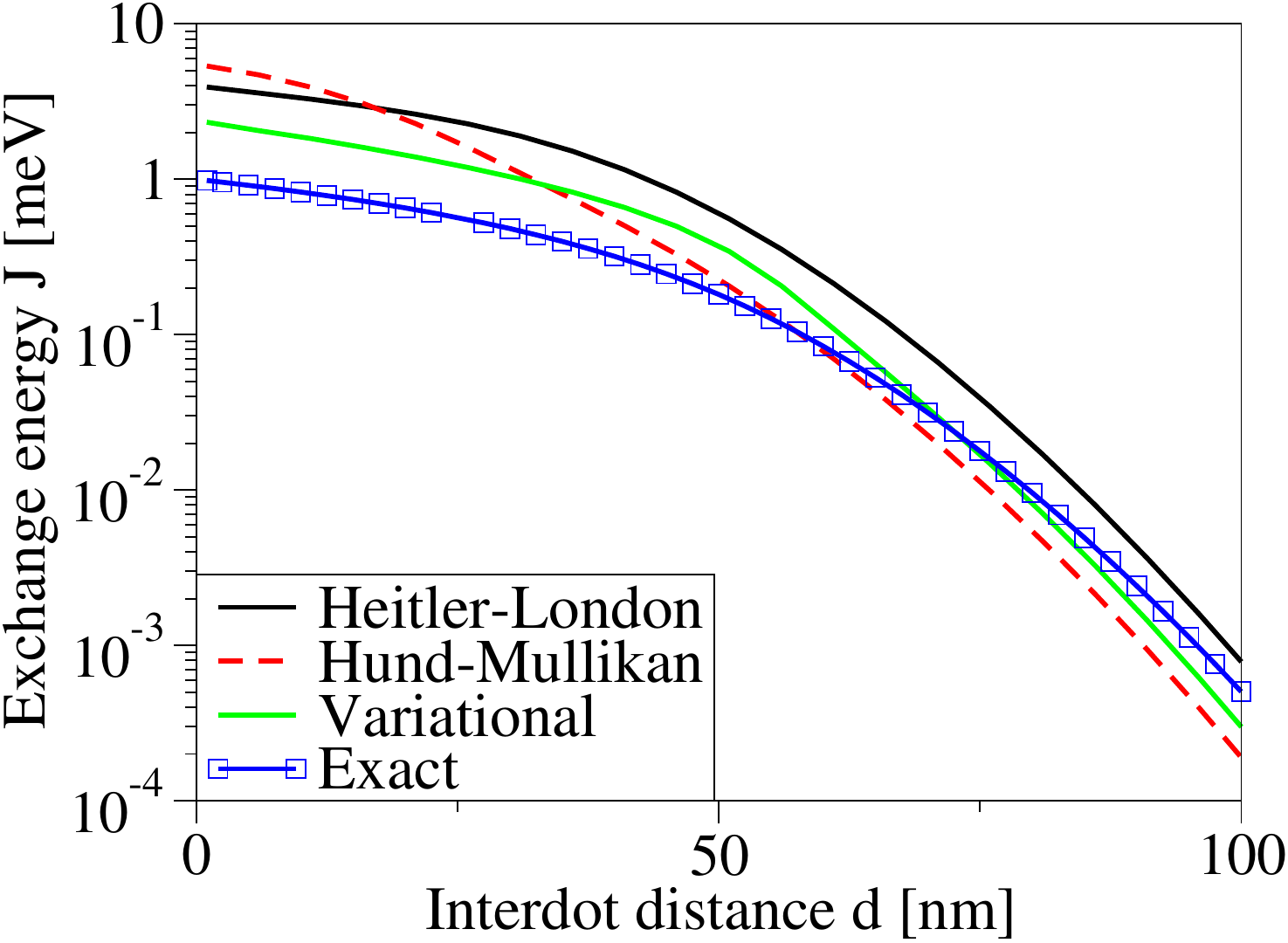}
 % JST0hw1.eps: 146421288x137584640 pixel, 300dpi, 1239700.25x1164883.25 cm, bb=(atend)
 \caption{{\small (Color online) Exchange energy as a function of the interdot distance  for two electrons in a double dot in a zero magnetic field.}}
 \label{fig:1}
\end{figure}
\section{Results}
\label{sec.IV}
Here we present results of the numerical calculations for the exchange energy using the methods listed in the previous.
In Fig.~\ref{fig:1} the exchange energy is plotted as a function of the interdot distance  for two electrons in a
 double dot in zero magnetic field. 
We observe that for large interdot distance the exchange energy falls off exponentially, a fact that all methods reflect correctly.  
This suggests that, at least in principle, an efficient control of the exchange energy can be achieved by increasing the potential barrier separating the dots and/or by increasing the interdot separation. At a small interdot distance all methods differ significantly from the exact result. We note here that, despite the common belief, the Hund-Mullikan
 (equivalent to Molecular-Orbital) method does not offer any improvement, even at small interdot distances (strong interdot couplings).
 The Variational method of the form that we choose, on the other hand, proves more robust, typically cutting the error of the Heitler-London to a half.
 \begin{figure}[h!]
 \centering
 \includegraphics[width=8cm,height=5.5cm]{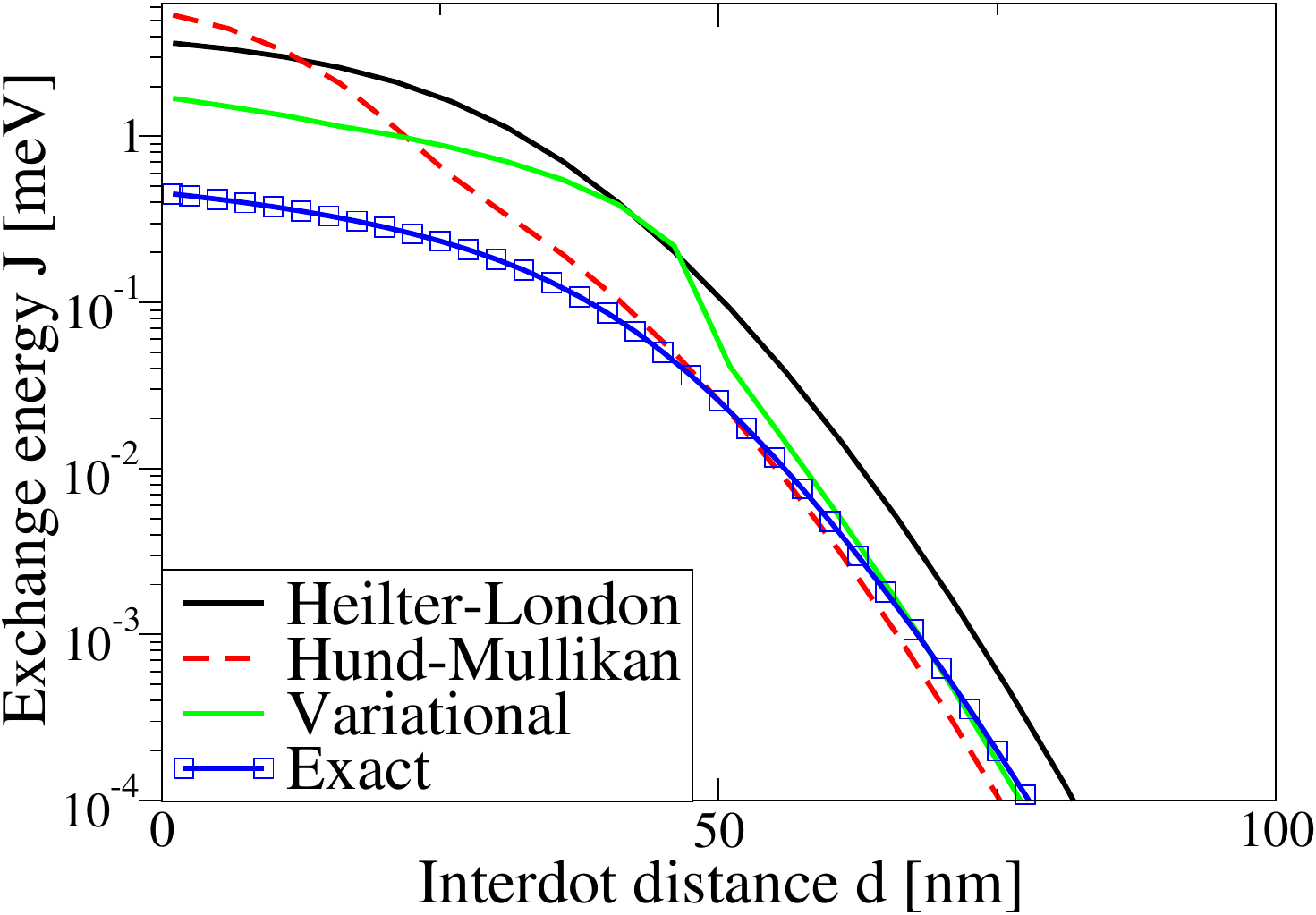}
 % JST0hw1.eps: 146421288x137584640 pixel, 300dpi, 1239700.25x1164883.25 cm, bb=(atend)
 \caption{{\small (Color online) Exchange energy as a function of the interdot distance  for two electrons in a double dot in a finite magnetic field of $1$ Tesla. 
}}
 \label{fig:2}
\end{figure}
 We will see on examples
 that follow, that these two features are generic and we explain them below.

 Figure \ref{fig:2} shows the exchange energy as a function of the interdot distance in a finite magnetic
field $1$T. The exchange energy decreases  faster than in zero magnetic field.
A simple explanation follows from noting that the natural length scale is the effective confinement length $\ell_{B}$, which 
drops with the magnetic field, as seen from Eq.~(\ref{eq:effe}).

\begin{figure}[h!]
 \centering
 \includegraphics[width=8cm,height=5.5cm]{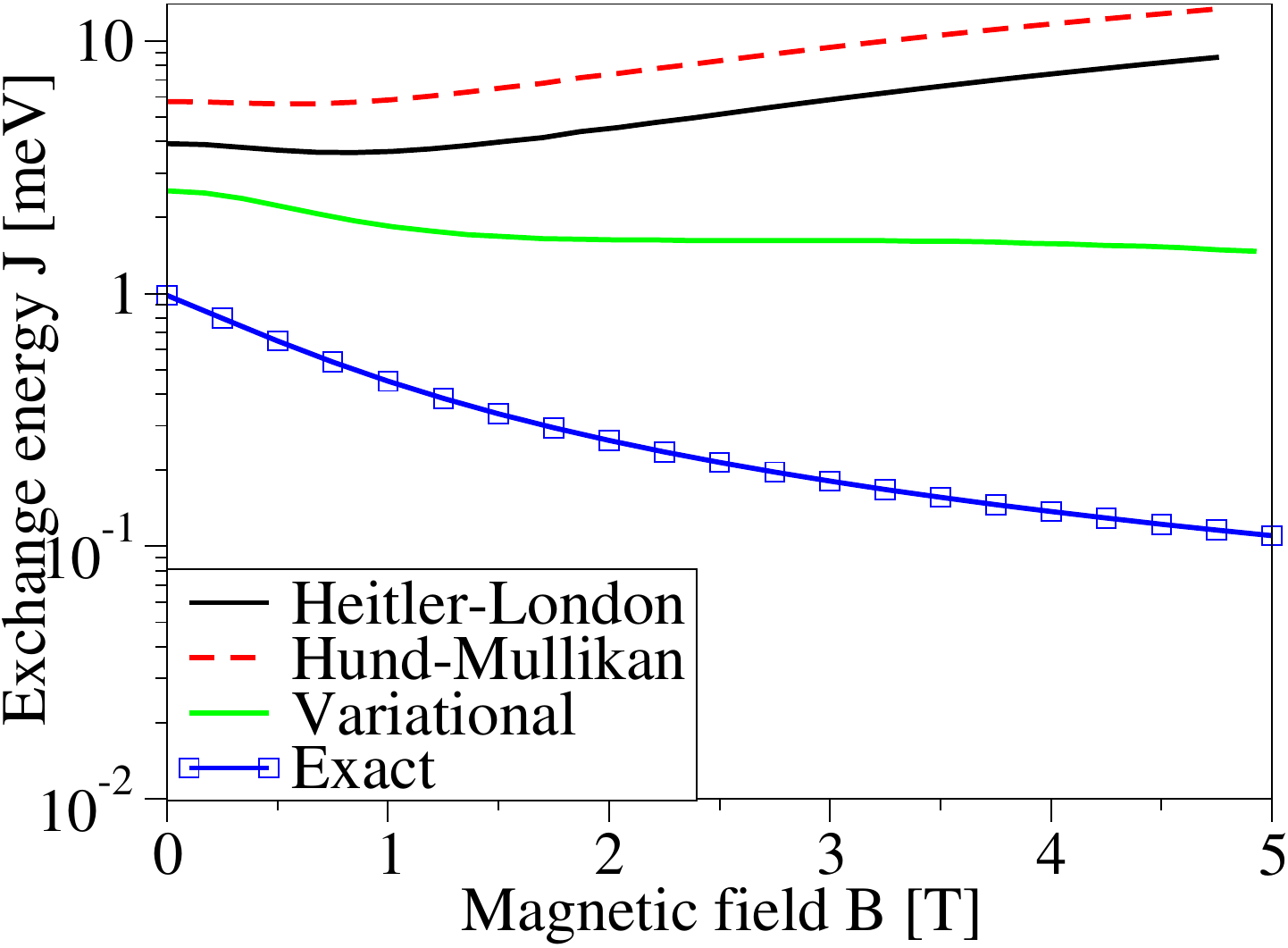}
 % JST0hw1.eps: 146421288x137584640 pixel, 300dpi, 1239700.25x1164883.25 cm, bb=(atend)
 \caption{{\small (Color online) Exchange energy as a function of the magnetic field  for two electrons in a single dot.}}
 \label{fig:3}
\end{figure}

Figure ~\ref{fig:3} shows the exchange energy as a function of the magnetic field for a
 single dot. We can see that as the magnetic field increases, the approximative methods become increasingly off the
exact results. Worse than that, except the variational method, even the trend is wrong (growth, instead of a fall-off)

\begin{figure}[h!]
 \centering
 \includegraphics[width=8cm,height=5.8cm]{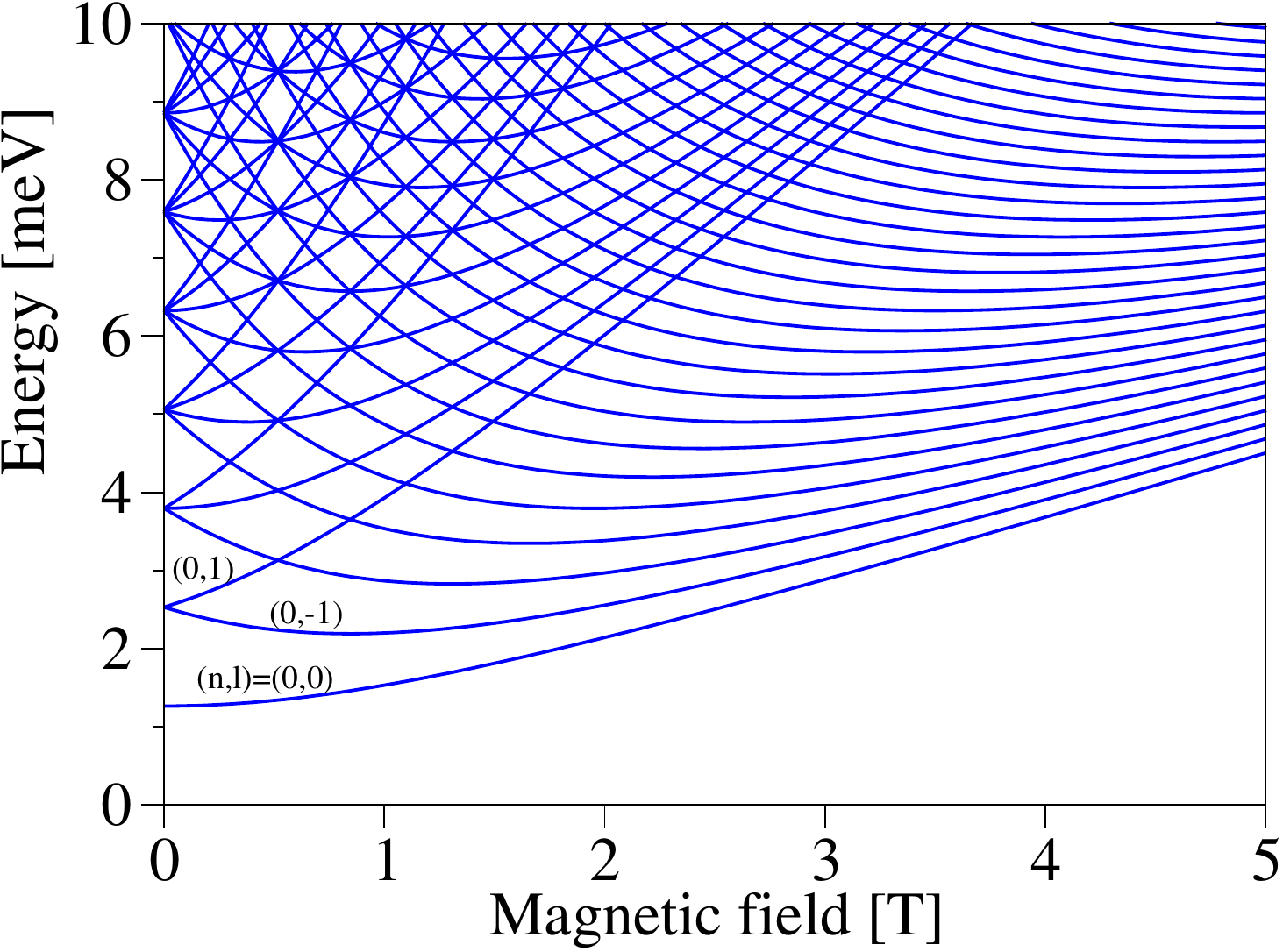}
 % JST0hw1.eps: 146421288x137584640 pixel, 300dpi, 1239700.25x1164883.25 cm, bb=(atend)
 \caption{{\small (Color online) Spectrum of a single electron in a single dot (Fock Darwin) as a function of the magnetic field.}}
 \label{fig:4}
\end{figure}

To understand this failure, we plot in Fig.~\ref{fig:4} the single electron single dot (Fock-Darwin) spectrum. One can appreciate the qualitative change
between the low and high field limit. In the first, the ground state is non-degenerate, while in the second, a highly degenerate Landau level forms. 
The
higher degeneracy, the more mixing and therefore worse results are expected for all methods based on a basis built from just a few single electron 
states. In other words, the effective strength of the Coulomb interaction grows with diminishing
of the energy separation of the single electron states due to the magnetic field \cite{PhysRevLett.65.108, PhysRevLett.67.1157}.

\begin{figure}[h!]
 \centering
 \includegraphics[width=8cm,height=5.5cm]{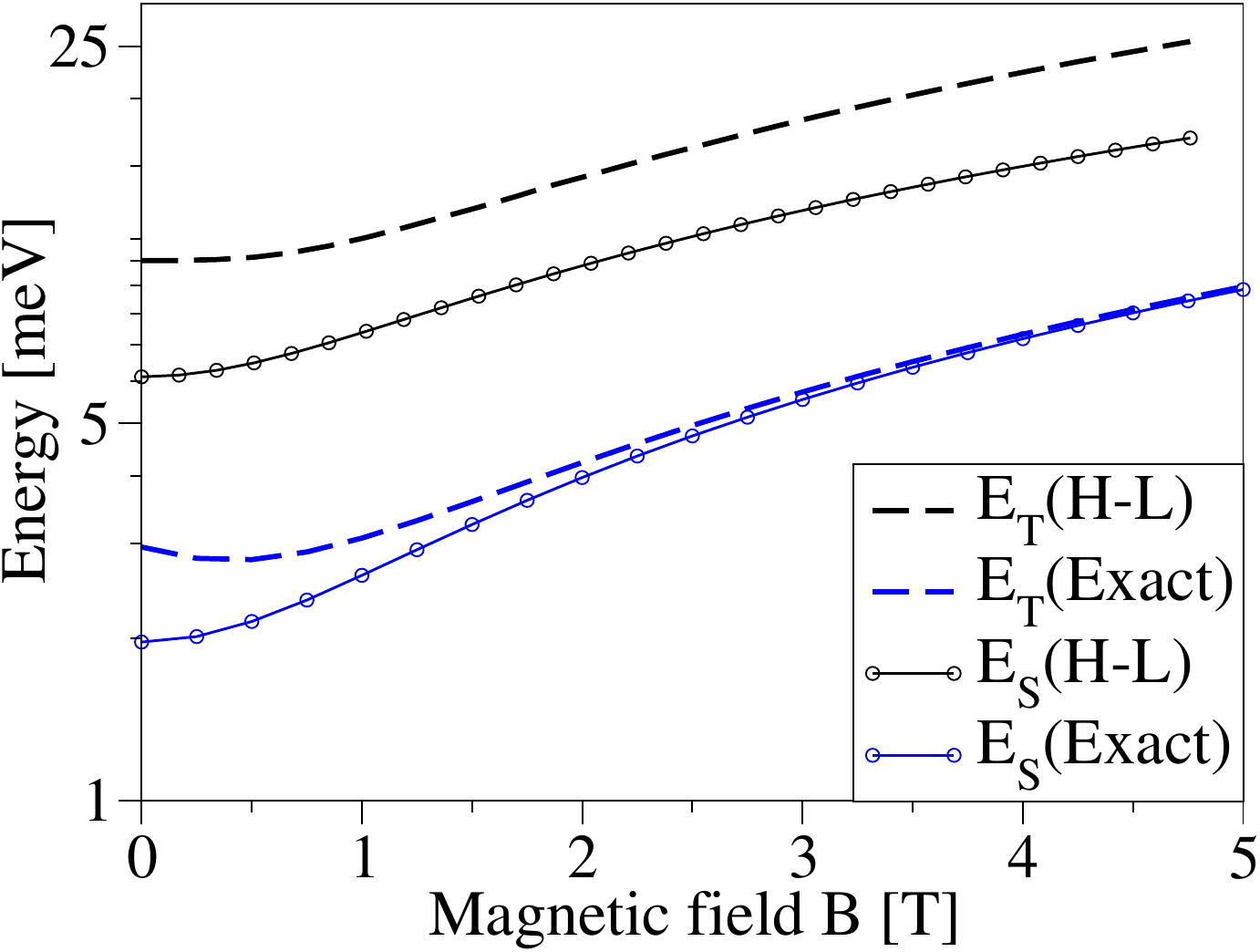}
 % JST0hw1.eps: 146421288x137584640 pixel, 300dpi, 1239700.25x1164883.25 cm, bb=(atend)
 \caption{{\small (Color online) Singlet and triplet energies in the Heitler-London method with their exact counterparts for a single dot. }}
 \label{fig:5}
\end{figure}

To gain further insight, we plot in  Fig.~\ref{fig:5} a comparison of the singlet and triplet energies in the  Heitler-London approximation with their
exact counterparts as a function of the magnetic field for a  single dot. 
From this we can see that both the singlet and the triplet are in the Heitler-London approximation similarly off
the exact results (that is, the failure is not due to just one of them). If operating as a qubit, the double dot will be manipulated at large interdot distances. 
\begin{figure}[h!]
 \centering
 \includegraphics[width=8cm,height=5.5cm]{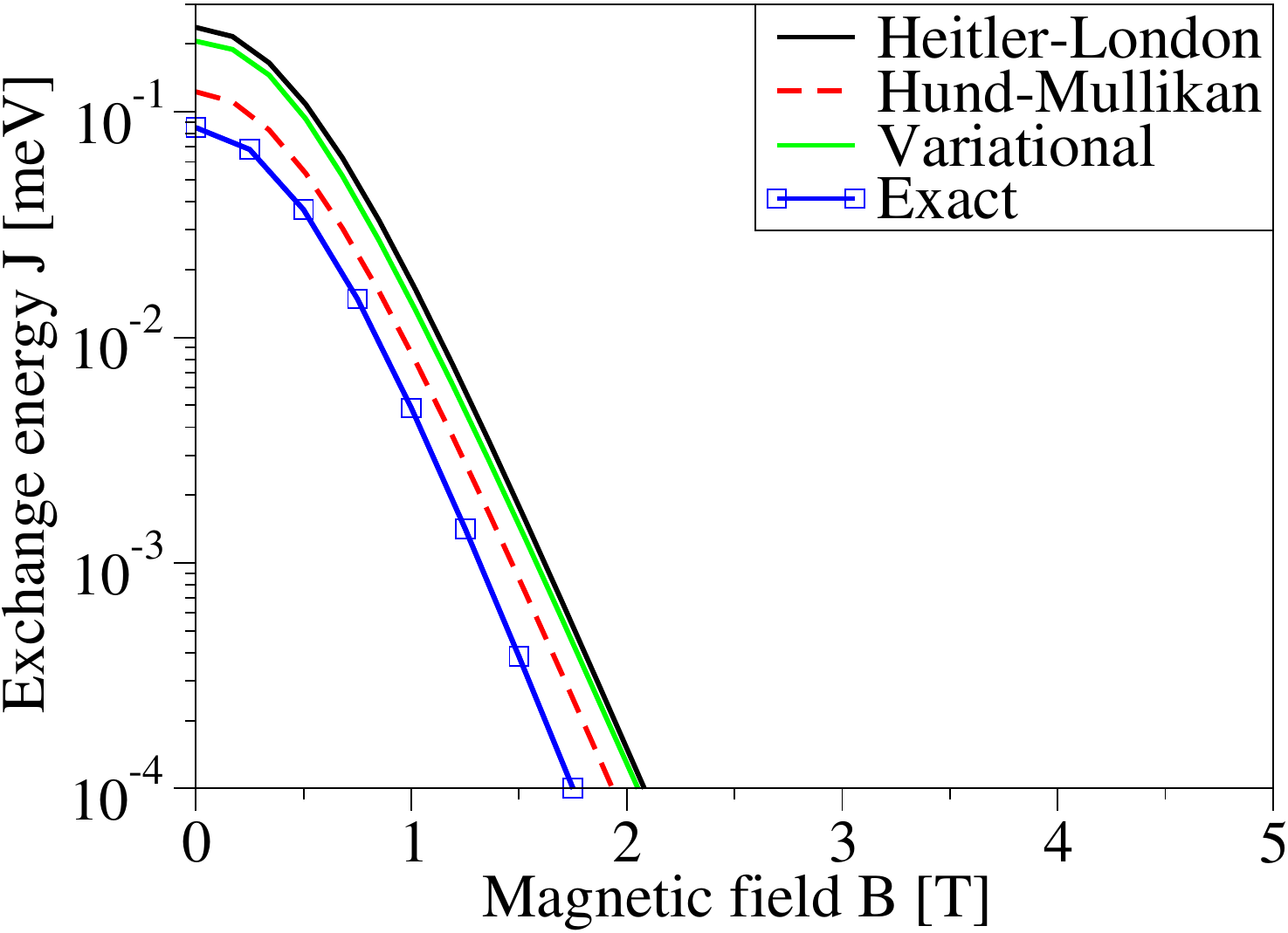}
 % JST0hw1.eps: 146421288x137584640 pixel, 300dpi, 1239700.25x1164883.25 cm, bb=(atend)
 \caption{{\small (Color online) Exchange energy as a function of the magnetic field  for two electrons in a two-electron double dot ($d=60$ nm).}}
 \label{fig:6}
\end{figure}

\begin{figure}[h!]
 \centering
 \includegraphics[width=8cm,height=5.5cm]{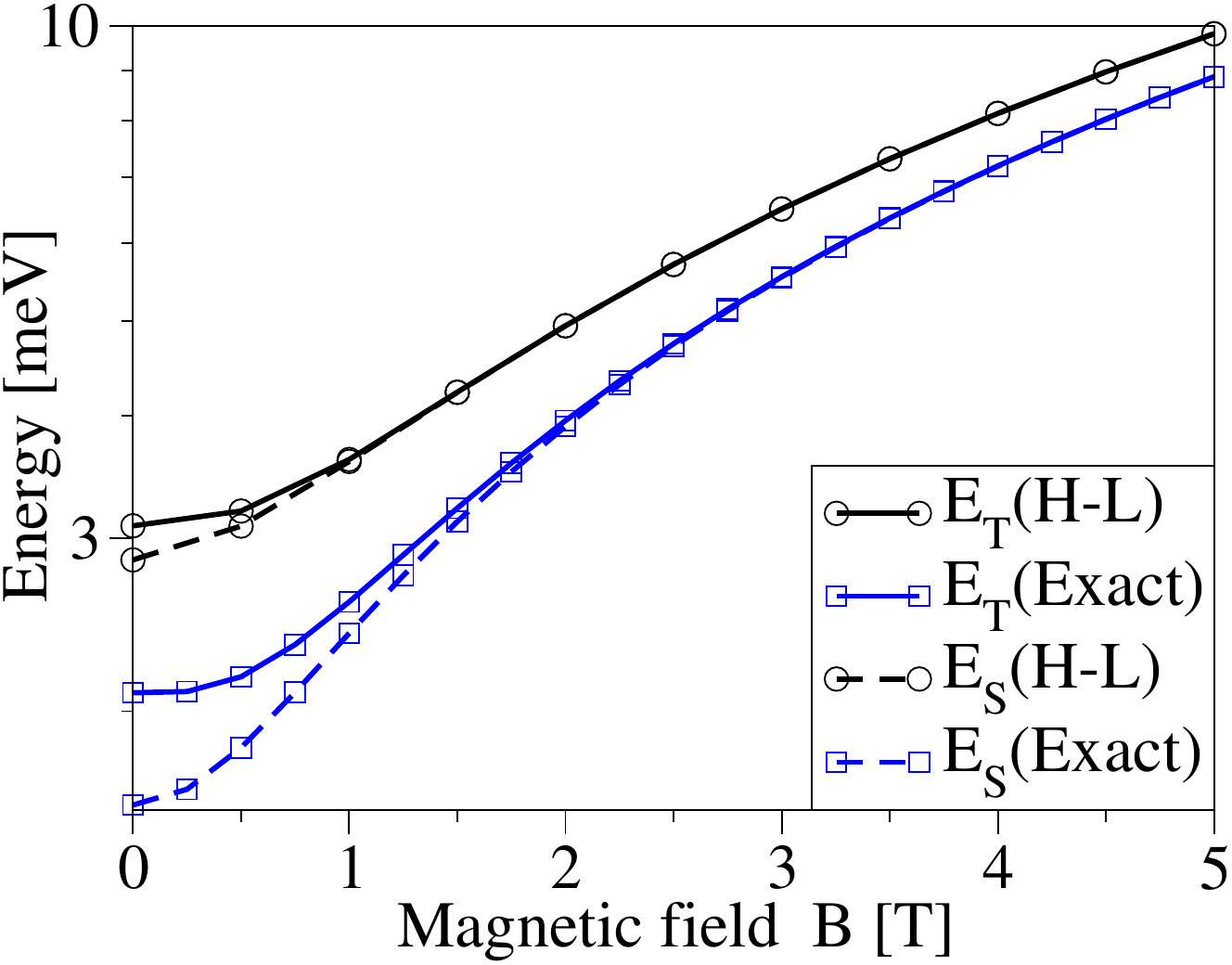}
 % JST0hw1.eps: 146421288x137584640 pixel, 300dpi, 1239700.25x1164883.25 cm, bb=(atend)
 \caption{{\small (Color online) Singlet and triplet energies in the Heitler-London method with their exact counterparts for double dot ($d=60$ nm).}}
 \label{fig:7}
\end{figure}
Figure ~\ref{fig:6} shows the exchange energy in this regime.
Surprisingly, the approximate methods reflect the exponential suppression of the exchange energy with the magnetic field correctly. This can be traced down to the fact
that in the Heitler-London method, the exchange energy is proportional to the overlap of the localized single electron states, from Eq.~(\ref{eq:103}). This
is illustrated further in Fig.~\ref{fig:7}, where we plot the singlet and triplet energies in the Heitler-London approximation together with their exact values in 
this regime. One can see that even though the approximate values do not converge to the exact one, their errors tend to compensate. As a result, the exchange energy falls towards
zero as the magnetic field is enlarged.
This finding also helps to understand the previously seen trends. Namely, as seen from Eq.~(\ref{eq:molor}), the Molecular-Orbital approximation expands, 
compared to the Heitler-London, the singlet subspace only. In the resulting energy pair, the error of the singled is reduced, while the triplet is untouched. Since the
exchange energy is the difference of the two energies, using the Hund-Mullikan/Molecular-Orbital is in fact detrimental compared to the Heitler-London 
method. The Variational method that we choose, on the other hand, treats the symmetric and antisymmetric wave-functions similarly. 
It is then natural to expect that their errors tend to compensate, resulting in more precise value for the exchange energy.

\section{Conclusions}
\label{concl}
To conclude, we studied the exchange energy in two electron single and double lateral quantum dots.
 We compared four approximate methods: Heitler-London, Hund-Mullikan, Molecular Orbital and Variational, 
with numerically exact results (the configuration interaction method).
We find that, compared to the much simpler Heitler-London method, the Hund-Mullikan and Molecular Orbital methods do not offer any improvement, 
whether at large or small interdot distances. We explain that noting the former two methods treat the singlet and triplet wave-functions differently 
leading to uncompensated errors. 
On the other hand the variational ansatz proves robust. At finite magnetic field, all approximate methods fail. This is a consequence of the qualitative
change of the single electron ground state. Finally, and most important, we find that all the approximate methods we study are free from the failure in the 
double dot regime.

\section{Acknowledgements}
We would like to thank Peter Sta\v{n}o for proposing the project and for his guidance to finish it.
This work was supported by the Ministry of Education, Science, Research and Sport of the Slovak Republic. 

\bibliography{biblio}
%\clearpage
%\appendix
%\include{Appendix/Appendix1}
%\section*{Supplementary Material}
%\label{supplement}

%\section{The SNS junction model}
%\label{appA0}
\end{document}